\documentclass[reviewcopy]{elsart}
\usepackage{amssymb}
\usepackage{graphicx}

\begin{document}
\begin{frontmatter}

\title{Interrelating Binarity, Physical Properties and
Orbital Evolution of Near Earth Asteroids: A New Numerical
Approach}

\author{Mihailo \v{C}ubrovi\'{c}}
\address{Department of Astronomy -- Petnica
Science Center, P. O. B. 6, 14104, Valjevo, Serbia and Montenegro}

36 pages, 3 figures and 1 table

\newpage
Proposed running head: Physical and Orbital Evolution of Binary
NEA

Send proofs to: Mihailo \v{C}ubrovi\'{c}

E-mail: cygnus@EUnet.yu

\newpage
\begin{abstract}
Possible connections between the physical properties of Near-Earth
Asteroids (NEA) and their orbital evolution were explored, with
emphasis on binary asteroids. Our main starting hypothesis,
suggested from the observations, was that the NEA population
contains considerably higher percent of binaries compared to the
Main Belt. The simulation covered simultaneously both orbital
evolution and evolution of physical properties, which is the main
improvement in comparison to the previous researches of this kind.
Physical evolution due to tidal forces and collision/disruption
events was described with TREESPH hydrocode. The results show
typical evolution paths of some types of NEA: asteroids with a
satellite, synchronized binaries, single fast rotators, double
asteroids. The two latter types and their physical properties
match closely to some observed NEA, while the two former ones have
not been detected in observational practice. We conjecture that
this is partly due to selection effects of current observational
techniques but partly also because some of them are likely to
evolve into different types of objects (e. g. single slow
rotators) through the processes we have not taken into account in
our simulation. The total percentage of binaries among NEA can be
estimated to 12\%. We have to notice, however, that more detailed
simulations, performed for longer time spans, are needed to
confirm our results.
\end{abstract}

\end{frontmatter}

\newpage
\section{Introduction}

The population of Near-Earth Asteroids (NEA) is generally
considered to be an especially interesting group of asteroids, in
part because of their complicated origin, and in part because of
their accessibility to observations. Physical properties and
physical evolution of these objects are relatively poorly
known\footnote{ Under ``physical properties'' we assume any
properties which are not orbital elements and are not directly
dependent upon the orbital motion, e. g. shape, binarity, rotation
state, internal structure, existence of precession, etc.}. The
basic concept of most models is the rubble-pile structure: an
aggregate of loosely (exclusively gravitationally) bond blocks.
Observational evidences for the rubble-pile structure (as well as
for other possible aggregate structures) can be found e. g. in
\cite{ast3agg}. The algorithm for simulations of this kind can be
based either on an N-body code \cite{zoe,rich98,solhil} or on a
hydrocode \cite{lovahrens,rich01,rich96}. Some recent papers take
into consideration both gravity and cohesive forces \cite{michel}.
The initial conditions of events like collision or close approach
to the Earth are usually estimated from coarse (and occasionally
quite non-physical) empirical distributions. Therefore, despite
some progress in understanding tidal and collisional processes
among the NEA, we still lack a consistent model of their physical
evolution and formation of binary systems. The laboratory
experiments of collisional processes can also yield some
information but their validity is questioned by the problems of
scaling.

On the other hand, the orbital evolution of NEA is known in much
more detail. A particular form of chaos is especially notable in
this population: although it is strongly affected by numerous
close approaches, the whole population seems to remain in an
approximately steady state. Despite the lack of long-scale
predictability for particular objects, some detailed models of NEA
population have been published \cite{rabin,neo}. According to the
mentioned papers, the $3:1$ and $\nu_6$ resonances seem to be the
most important source of NEA although the Mars crosser (MC)
population also has some importance.

In this paper, we give some results of our numerical simulation of
orbital and physical evolution of NEA, with special emphasis on
the formation and properties of binary systems. The main
contribution of this paper is the simultaneous treatment of
orbital motion and collision/disruption events, which is, as far
as we know, carried out for the first time. One of the primary
objectives of this research is the explanation of observational
evidences that the NEA population has substantially larger percent
of binary asteroids than the Main Belt \cite{margot}. Tidal
deformations (and disruptions) during close approaches, collisions
among the NEA, as well as mutual tidal influences in binary
systems, cause very complicated perturbations of both orbital
elements and physical properties. Many of the mentioned processes
are probably interrelated. Therefore, we expect this simultaneous
treatment to resolve some previously unknown mechanisms and to
explain evolution paths of some typical NEA objects. Some of our
findings are not entirely new but at least yield a direct
confirmation of earlier proposed scenarios, while others contrast
to the usual understanding of binaries formation.

Of course, the most significant drawback of this approach is the
large amount of calculations that have to be done since many
processes are being simulated at the same time. This limits the
number of objects involved as well as the interval of the
simulation. An important factor is also the limited resolution of
the hydrocode. In addition, some of the initial conditions are not
known very well, which also affects the validity of the results.
Therefore, this research should only be considered as a
qualitative picture of some processes among the NEA, given on a
fictitious representative ensemble of objects -- it is not a model
of NEA population nor does it give valid predictions for the
evolution of any particular object.

The second section gives a short review of the methodology,
including the orbital integrator, the treatment of
non-gravitational influences, the dynamical model of binary
systems and the hydrodynamical model adopted for the description
of physical evolution. The third section describes the
implementation of the simulation and the simulated system. Results
are given in the fourth section. Some theoretical implications of
the results, as well as some comparisons to the previous research
are discussed in the fifth section. The sixth section sums up the
conclusions.

\section{Methodology} \subsection{Orbital integrator}

Basic request for the orbital integrator was efficiency rather
than precision. As it is mentioned in the first section, we were
seeking only for a qualitative description of the orbital
movement, not for the exact orbital elements. We adopted a
second-order symplectic integrator, proposed by \cite{symp},
usually known as the ``Mixed-Variable Symplectic'' integrator
(MVS). Its efficiency is usually considerably higher than that of
ordinary integrators (e. g. Bulirsch-Stoer), while the precision
remains valid for many purposes, despite the (typically) low order
of MVS integrators.

In our integrator, we used a second-order algorithm. The
integration included four planets (Earth, Mars, Jupiter, and
Saturn). The Mercury's mass was added to that of the Sun, while
the Moon's mass was added to the Earth's mass. Asteroids were
included as test particles.

Our integration scheme is, however, somewhat different from the
original concept of \cite{symp}, since it includes some
improvements, proposed by \cite{symp92,symp94}. This version of
the MVS integrator gives the error of the order
$\mathcal{O}\left({\varepsilon^2\tau^2}\right)$, where
$\varepsilon $ denotes the planets to Sun mass ratio, and $\tau$
stands for the average time step.

The non-gravitational forces (the Poynting-Robertson drag and the
Yarkovsky force) could not be included directly since they make it
impossible to write the equations of motion in Hmiltonian form
needed for our integrator. As it is well known, these effects are
largely negligible for objects that exceed about 100 m. However,
some recent results \cite[and references therein]{yark} show that
the Yarkovsky effect could significantly influence evolution of
some objects by putting them into mean motion resonances.
Therefore, we decided to include this effect by applying
periodical corrections to the semimajor axes of the asteroids'
orbits, using the linear approximation derived by \cite{yark}. His
equations do not separate the seasonal part (dependent on the
orbital period) and the diurnal part (dependent on the rotation
period) of the effect but treat them together. We omit them here
for the sake of conciseness; they are very complicated and can be
found in the aforementioned reference. It is enough to mention
that the only parameters that correspond to particular asteroid
are density, mean radius, thermal capacity, and thermal
conductivity. The former two characterized each asteroid in the
simulation (for details see the third section) while the latter
two were fixed, calculated from the parameters of the equation of
state (see the third subsection of this section).

\subsection{Dynamical model of binary systems}

As it is well known, binary systems undergo mutual tidal
perturbations, which affect their motion around the mutual center
of mass. Close approaches to the Earth are also expected to change
their dynamical nature. A simple dynamical model was used to
describe these events, based on some analytical and
semi-analytical results.

Dynamic state of each binary system was characterized by spin
vectors, eccentricity and semimajor axes of the components (the
latter two, together with the masses, determine the orbital
frequency around the center of mass)\footnote{ We mention here
only those parameters which are included in the dynamical model of
binary systems; others are mentioned in the next subsection, and
all the parameters are listed in the third section.}. The tidal
influence was calculated as a series of periodical changes in
orbital frequency, eccentricity, and rotational periods, using the
analytical expressions of classical tide theory \cite{sat}. Since
these expressions are complicated and can be found in many
classical references, we shall, as for the Yarkovsky drift, omit
them.

The close approaches were treated in a semi-analytical way,
following the idea of \cite{alpha}. The relative changes of energy
and angular momentum of the system can be expressed as:

\begin{equation}
\label{eq1} \frac{\Delta E}{E}=\frac{G^2 m_A
m_P}{Vb^2}\left(-2E\right)^{-\frac{3}{2}}I
\end{equation}

\begin{equation}
\label{eq2} \frac{\Delta L}{L}=\frac{1}{2}\frac{\Delta E}{E}
\end{equation}

\noindent where $m_P$ and $m_A$ denote masses of the planet (in
this case Earth) and the asteroid (the whole binary system), while
$V$ and $b$ represent the velocity of the planet in the reference
frame of the asteroid and the impact parameter, respectively. The
dependence of the perturbation upon the geometry of the approach
(parameterized with three angles, see \cite{alpha} for details) is
contained in the non-dimensional integral $I$. Analytical solution
of this integral does not exist in the general case. \cite{alpha}
approximate it with a Gaussian random variable, starting from the
rectilinear approximation. We, however, calculated (numerically)
and tabulated the value of $I$ for different geometries of the
encounter, which allowed us to adopt a more realistic, hyperbolic
approximation. During the simulation, the actual value for each
close approach was calculated as a linear interpolation from the
table with respect to the independent parameters, which allowed
better efficiency than immediate integration.

\subsection{The hydrodynamical model of asteroids}

The core of our hydrodynamical model is the hydrocode usually
known as TREESPH, given by \cite{treesph}. It is a combination of
the Smoothed Particle Hydrodynamics code (SPH), and the
Hierarchical Tree Method (HTM), an algorithm developed for
hierarchical work with clusters of objects. The latter gives to
TREESPH also some good features of N body algorithms. All the
details of our model and its specific properties are not of
interest here and will be published elsewhere. In this subsection,
we shall mention only some basic ideas and differences from the
published models.

First, we shall briefly revisit the SPH formalism. This method was
given by Gingold and Monaghan (1977; according to \cite{treesph})
and it has become widely accepted for applications that require
high efficiency and modest precision. It is a Lagrangian particle
code, which is mathematically based on integral interpolation. The
interactions among the particles are described via an
interpolation function. Value of a physical field $f(\vec{r})$ in
a given location (i. e. for a given particle) is calculated as:

\begin{equation}
\label{eq3} f_{SPH}\left(\vec{r}\right)=\int
W\left(\vec{r}-\vec{r'};h\left(\vec{r},\vec{r'}
\right)\right)f\left(\vec{r}\right)\d\vec{r}
\end{equation}

\noindent where $W$ denotes the kernel, and the integration is
performed over the whole volume of the system (in this case
asteroid). The $h$ parameter is the smoothing length, which
roughly corresponds to the resolution of the hydrocode. The kernel
characterizes the strength of the interaction, so it typically
drops very quickly when $r-r'$ becomes large. Its integral is
normalized to unity. As it can be seen from (\ref{eq3}), the
smoothing length is spatially variable, and it changes both
locally and globally. This spatially adaptive smoothing length is
a peculiarity of TREESPH; the original SPH uses constant smoothing
length. In practice, of course, the integration turns into
summation over a set of discrete particles. For the kernel, we
adopted the cubic spline proposed by Monaghan and Lattanzio and
given by \cite{treesph}. Its value drops to zero for $r-r'>2h$.
The evolution of the system is described, as usually, by
continuity equation, momentum equation, energy equation and
equation of state (EOS). The procedure for discretization of these
equations can be found in the aforementioned reference and will be
omitted here. The well-known fourth-order adaptive step
Runge-Kutta algorithm \cite{nr} was used for the integration.

For the EOS were adopted the Tillotson equations \cite{benz}.
Since they are somewhat less known, we will describe them here.
The basic idea is to consider analytically only two extreme cases
(concerning the energy of the system); otherwise, the resulting
EOS is calculated via linear interpolation. If the volume density
of energy is less then the energy of incipient vaporization
($E<E_{iv}$) the pressure is given by:

\begin{equation}
\label{eq4} P=\left[a+\frac{b}{1+\frac{E}{E_0\eta^2}}\right]\rho
E+A\mu +B\mu^2
\end{equation}

\noindent If the volume density of energy grows larger than the
energy of complete vaporization ($E>E_{cv}$) the previous equation
becomes:

\begin{equation}
\label{eq5} P=a\rho E+\left[\frac{b\rho
E}{1+\frac{E}{E_0\eta^2}}+A\mu\exp\left(\beta-\beta\frac{\rho_0}{\rho}\right)\right]\exp\left[-\alpha\left(\frac{\rho}{\rho_0}-1\right)^2\right]
\end{equation}

\noindent The parameters $A$, $B$, $a$, $b$, $\alpha$, $\beta$,
$E_0$, $E_{iv}$, $E_{cv}$, $\rho_0$, $\mu$ are dependent on the
material. Following  \cite{lovahrens}, we adopted their values for
basalt taken from \cite{benz}. Density of each particle was also
taken to be equal to the density of basalt ($\rho=2.7$ g cm$^-3$).
Initially, the particles were distributed in a face-centered cubic
array, which gives the average density about $1.8$ g cm$^{-3}$
(less than the theoretical value because of finite dimensions of
asteroids).

The role of HTM is to make the calculations of interactions among
the particles more efficient. Summation of interactions between
each two particles in general case requires an
$\mathcal{O}\left(N^2\right)$ algorithm. In HTM, particles are
formally arranged in clusters, which may replace single particles
in hydrodynamical calculations. During the force evaluation, the
algorithm creates a tree with clusters (which have the physical
meaning of three-dimensional cells in space) at the nodes.
Therefore, distant particles contribute to the resulting force
only as clusters while near ones contribute to it directly. Of
course, some empirical criterion for clustering has to be adopted;
we have used a polynomial law derived from numerical experiments.
The whole procedure (tree construction and the force evaluation)
is now performed in $\mathcal{O}\left(N\log N\right)$ time, and it
makes the model plausible also for the ballistic phase of an
impact (unlike the traditional SPH code).

As it has already become clear, the model asteroid is simply a set
of gravitationally interacting particles in an external
gravitational field (originated from some other body, see next
section for details). We completely neglect material strength, so
this is a purely ``rubble-pile'' model. Friction and fractures are
also neglected. Particles are represented as non-elastic spheres
of finite radius. The reference frame is always the center of mass
of the system. We used this model to simulate collisions and tidal
disruptions.

Criterion for escaping particles was one of the most problematic
issues since the safest solution -- direct integration long enough
for all the escaping particles to actually escape -- was not
possible due to computational reasons. Speed criterions are not
plausible because of the non-sphericity of the asteroid. We
implemented a criterion suggested by W. F. Bottke (e-mail
communication): simply to look for the particles with the absolute
value of potential energy larger than the kinetic energy. Although
somewhat dangerous because of artificial energy oscillations,
which are sometimes produced by SPH, this criterion generally
seemed realistic. The second major issue was the detection of
satellites that may form during a collision/disruption event.
Since the satellites of asteroids are among the main objectives of
our research this criterion had to be imposed more exactly, also
because we had to calculate its starting orbital elements (to be
used by the dynamical model, see the previous subsection). We
decided to use the semianalytical results based on Hill's
equations \cite{has}, which allowed us to calculate the orbital
elements and mark the fragment as a satellite if its orbit turns
out to be stable (i. e. orbit which does not include the collision
with the main body; hyperbolic orbits are eliminated by the
previous test, since they are equivalent to the escape). Finally,
the simulation of a particular event was considered complete when
all the non-escaping particles either fall back to the asteroid,
or start moving on elliptical orbits as satellites.

\section{Model and simulation}

The simulated system was designed as a representative ensemble of
objects that are in source regions for NEA population and
therefore can be expected to become NEA relatively quickly. Since
we supposed that most of the interesting events (e. g. collisions)
happen during the transit to the NEA region, we decided to start
with objects in the source regions, and not with objects which
have already become NEA. We also decided to work with fictitious
objects, since any selection of known objects could result in a
biased ensemble and, on the other side, as we have already
mentioned, the precision of the simulation does not give valid
predictions for any particular object. Therefore, our simulation
was intended to follow the migration processes in general and to
allow the analysis of typical NEA after they pass through the
transition mechanisms and, possibly, collision/disruption events.

The simulated system contained 160 objects -- one sixth of the
estimated current NEA population \cite{neo}. The interval of the
simulation was 10 Myr. The initial orbital elements were
calculated from distributions given in the aforementioned
reference. The mentioned paper considers the following source
regions: the $3:1$ resonance, the $\nu_6$ resonance, the MC
population, the comets of Jupiter family (JFC), and the outer belt
(OB). We omitted the last two sources (from which come 14{\%} of
NEA, according to \cite{neo}) since their transition mechanisms
tend to be very complicated and beyond the scope of our research.
Relative populations of $3:1$, $\nu_6$ and MC regions were,
respectively 44$\%$, 29$\%$, and 27$\%$ (the mentioned 14$\%$ of
JFC and OB were proportionally distributed among the first three
regions). For exact distributions of orbital elements, see
\cite{neo}.

Each object was characterized by orbital elements, mass, spin
vector and hydrodynamical model (i. e. coordinates and momenta of
each particle, which determine also the density, dimensions and
shape). To make the comparison with the observational data easier,
we also introduced the mean radius. All the binary systems were
marked with a flag since they required two additional parameters
-- semimajor axis and eccentricity of the orbit around the mutual
center of mass. We supposed that the starting population contained
no binaries.

The starting distribution of spin vectors (concerning both
orientation and intensity) is generally an unexplored subject.
Most authors \cite{chau,yark} assume an isotropic distribution of
spin vectors, so we followed them, largely to make comparison with
\cite{chau} easier. Possible problematic consequences of this
decision are discussed in the fifth section. For the periods of
rotation, we adopted the Maxwellian distribution obtained for Main
Belt objects e. g. by \cite{pravhar}. Size distribution
(distribution of radiuses) was taken from \cite{gomes}; this is an
exponential distribution, a widely accepted form for various
objects. Finally, the starting shapes were triaxial ellipsoids,
with axial ratios distribution taken from observational data given
in Uppsala Photometric Catalogue of Asteroids,
\textit{http://www.astro.uu.se/~classe/projects/apceng.html}.

The organizational base of the simulation was the orbit
integrator, which was programmed to ``turn on'' the hydrodynamical
simulator if a collision/disruption event is likely to happen. At
the end of each time step, a test was performed to check if the
asteroid enters the sphere of influence of some other object (an
asteroid or a planet). If a direct collision with a planet
happened, the asteroid was discarded from the simulation.
Otherwise, the hydrocode was activated, which performed the
calculations during the collision or close approach (the latter
results in deformation and, in the extreme case, disruption). The
Yarkovsky drift and, for binaries, the tidal drift were added
periodically, as it was described in the previous section. If an
asteroid breaks into fragments, at the end of the hydrodynamical
simulation each fragment becomes a separate asteroid and continues
its evolution separately; its spin vector is calculated from the
equation of angular momentum, taking into account also the angular
momentum carried by small debris. Bodies with mean radius less
than $100$ m were discarded since such small bodies require a much
more detailed treatment of cohesive forces and non-gravitational
influences. Objects that cross the Jupiter's orbit were also
discarded.

The implementation of the simulation cannot treat encounters of
three or more bodies nor can it treat more than one
collision/disruption event at the same time step; bearing in mind
the probability of these events, we did not take this for a
serious disadvantage.

\section{Results} \subsection{General remarks}

It is clear that a simulation with so many parameters gives a
large amount of numerical results; their detailed analysis is not
of interest for this research (although it could be interesting in
general). We shall focus only on some characteristic results,
which are important for the objective of this paper.

Evolution of the simulated system generally corresponds with
current theoretical knowledge about NEA migration and evolution.
Processes of transition follow the usual path; the most efficient
mechanisms are, as expected, close encounters of the planets, and
the most efficient source is, again expected, the $3:1$ resonance.
The MC region was somewhat more efficient than expected. After
about 2 Myr the system became relatively stable and the number of
bodies in the NEA region became nearly constant. Most of the
particles survived until the end of the integration, despite
largely chaotic nature of their evolution.

Collisions of asteroids also seem to fit well into current models.
The outcome depends on the mass ratio and impact angle, while the
relative speed tends to be less important. It seems that the
reaccumulation of collisional fragments has a more prominent role
than in previous researches \cite{zoe}, which may be a consequence
of partially N-body nature of TREESPH. The bottom size limit for
formation of stable rubble-pile objects seems to be about 100 m --
200 m. Of course, these remarks should be treated carefully, as
they are only our general notes about the collisions; they are not
a result of systematic analysis.

The tidal forces act relatively slowly but in long intervals they can become
key factors for an object's evolution. Low-speed approaches tend to be the
most efficient disruption mechanisms while the fast ones usually only
slightly deform the asteroid. Still, the outcome largely depends on the
initial physical properties of an asteroid. Only in two cases, we detected a
complete disruption into many fragments.

Before we start the analysis of the particular types of detected
binaries, we shall just briefly comment the spin rate
distribution. Among the numbered NEA, this distribution is more or
less uniform \cite{pravhar} in contrast to the Maxwellian
distribution of the Main Belt objects. The natural explanation
would be that the NEA do not have the time to achieve a steady,
collisionally-relaxed state due to their short lifetime. We
conjectured that binary systems are subjected to the more intense
evolution (collisional but also tidal) than the rest of NEA
population so we expected to find a distribution somewhat more
similar to the Maxwellian by taking only the binaries into
account. The histogram can be seen in Fig. 1. Strictly speaking,
in order to cheque the consistency with the Maxwellian
distribution, one would have to normalize the spin rates with
respect to the diameter \cite{pravhar} but this is not plausible
(nor very necessary) for a small set consisting of objects of
similar size, as it is the case here. The data for the real
objects (for Fig. 1 as well as for the rest of the paper) were
taken from Binary Near-Earth Asteroids page, maintained by Petr
Pravec and Peter Scheirich,
\textit{http://www.asu.cas.cz/~asteroid/binneas.htm}.

PLACE FIGURE 1 HERE.

The observed objects are all fast rotators, rotating probably on
the edge of breakup \cite{pravhar}. Our simulation shows the
similar peak of the fast rotators but also includes a number of
slowly rotating bodies, which makes the distribution visually
somewhat more similar to the Maxwellian but still far from it in
any quantitative (statistical) sense. Together with the
overabundance of the slow rotators among binaries, significant
lack of slowly rotating objects is found among the single ones.
This problem will be dealt with later; we just remark that it is
probably a combination of observational selection effects and
various effects not taken into account in this simulation.

\subsection{Formation and evolution of particular types of objects}

We shall first describe four types of "evolved" objects that could
be clearly distinguished at the end of simulation. Their most
important characteristics are summed up in Tab. \ref{tab1}. As
mentioned, given numerical values should be treated only as rough
estimates. Of course, besides these tidally/collisionally evolved
objects, there were many ``non-evolved'' objects -- single
asteroids with no peculiar features; we focus only on those, which
had been subject to intense evolution processes.

PLACE TABLE 1 HERE.

The first type in the table, double asteroids, denotes the binary
objects with non-synchronized rotation (revolution around the
mutual center of mass with period different from the rotation
period). Under synchronized binaries, we assume the systems that
rotate as a rigid body (with a unique spin rate for the whole
system); their components are usually very near but they can also
be well separated from each other. Because of that, we hesitate to
call them contact binaries, although most of them are very close
systems (in contact or nearly in contact). To the third type --
asteroids with a satellite -- belong objects with the mass ratio
less than $1/10$, i. e. systems in which one component clearly
dominates the other. Under single fast rotators we assume single
asteroids with period of rotation shorter than 4 hours. We did not
detect single slow rotators, which have been observed
\cite{pravhar}, though we did detect slowly rotating binaries,
usually among the contact systems (which have not been observed
thus far); we shall consider this problem in more detail later.
The double asteroids and the single fast rotators are well known
from observational practice \cite{margot,pravhar}, so we can say
we have reproduced some typical objects. Asteroids with a
satellite have not been observed among NEA, nor have been the
contact binaries (as we already remarked). Relative abundances of
the four types are given in Fig. 2.

PLACE FIGURE 2 HERE.

Double asteroids were detected mostly as outcomes of tidal
disruption. Of course, tidal effects influence primarily the Earth
crossers; however, these objects occasionally become ejected from
the region of the Earth crossers so we could detect them in the
whole NEA belt. Mutual tidal perturbations may lead to collision,
ejection, or formation of synchronized (usually contact) binaries.
However, many systems of this kind were remarkably stable and were
able to survive until the end of the simulation. As we already
pointed out, that these systems usually have very fast spin rates
(thus accounting for the peak of fast rotators in Fig. 1, which is
consistent with the observations \cite{margot}.

For the synchronized binaries, we have noticed two formation
scenarios. The first one is tidal evolution of the previous type,
when the components of a separated binary synchronize and come
close to each other. The second scenario is tidal disruption
during close encounters. In this case the components may
synchronize while staying well separated. Speed of motion during
the encounters is typically lower than in the previous case, so
the tidal forces become less efficient. Rotation of synchronized
binaries is usually slower than for single asteroids, which is
expected. This type of asteroids has not been observed, which can
be explained with the inability of light curve techniques (upon
which results are chosen also the objects for the radar analysis)
to detect synchronized systems \cite{ast3sat}. It could also
happen, however, that contact systems are likely to evolve into
some other objects (see the next section).

Asteroids with a satellite seem to be a typical outcome of
collisional evolution. According to our simulation, they can be
formed only in collisions. Satellite usually forms from ejected
material of both components. For colliding objects of similar
masses, the most probable outcome is a single fast rotator (the
next type) while for larger mass ratios the usual outcome is an
asteroid with a satellite. In extreme cases, the only outcome is
ejection of many small particles, which do not form a satellite.
It is interesting that collisions cannot produce (at least in our
simulation) separated binaries which appear as a natural
``transient'' form between synchronized binaries and asteroids
with satellites. However, we have to emphasize the instability of
asteroids with satellites. As it can be seen in Table 1, only
three such objects survived for a considerably long time.
Satellite usually becomes lost very soon after its formation. This
could be one of the reasons that these objects have not been
detected: their lifetime is very short. Another important reason
is the inability of current observational techniques to detect
small satellites \cite{ast3sat}: all such objects were detected
via either direct flyby of certain Main Belt asteroids.

Single fast rotators were also formed mostly during collisions if
the mass ratio of the colliding bodies is of the order of unity.
One object of this type was also formed as a consequence of
collision of components in a binary system. The latter case,
however, seems to be very improbable in practice, as it requires
very exact alignment of angular momenta, in order to produce a
large enough resulting momentum. During formation of fast
rotators, asteroids usually suffer significant mass loss, which is
certainly due to many small fragments that become ejected.

The overall percentage of binaries in the ensemble varied about
10$\%$--15$\%$. For the expected mean percentage we obtain 12$\%$,
which, according to \cite{neo}, gives about 110 binaries. This
agrees nicely with the analysis of observational data, which gives
the percentage of binaries about 16$\%$ \cite{ast3sat,margot}.

One of the most interesting aspects is, of course, the dynamical
evolution of binary systems. Fig. 3 shows the dependence of the
final rotational period upon the orbital elements of the binary
system -- eccentricity and semimajor axis (compare with figures in
\cite{chau}). The figure has been obtained by interpolating the
parameters of objects detected in the simulation. Therefore, it is
just a rough visualization of our results. Some parts of the
figure (especially those which are near the contact line, see the
figure caption for explanation) are rather uncertain estimates,
statistical in nature. Still, we think that the semimajor axis --
eccentricity -- period dependence is illustrative, both as giving
the general picture of the angular momentum distribution and as a
kind of confirmation (by the means of "direct simulation") of the
Monte Carlo results obtained by \cite{chau}.

PLACE FIGURE 3 HERE.

The general trend is slowdown of rotation with increase of
semimajor axis and eccentricity. A rather strong correlation with
the initial conditions can be seen. However, we could not detect
any significant influence of the spin vector orientation. This is
in clear contradiction with \cite{chau} which emphasize strong
instabilities of retrograde rotators. A possible cause is that in
our, hydrodynamical model asteroids gradually lose most of the
rotational energy on internal ``heating'' (random motion of
particles of the asteroid) so the additional tidal action which
appears in the case of retrograde rotation does not have
sufficient energy to cause the collision of components (which
happens in model of \cite{chau}). Other aspects of the dynamical
evolution of binary systems are qualitatively similar to the
results of the mentioned authors: close approaches and tidal
forces make asteroids lose energy, which causes either collision
or ejection of one component. In the latter case, the larger
component loses most of its angular momentum.

Data for the real objects are also shown in the figure.
Unfortunately, only seven objects have known (at least
tentatively) eccentricities, so this data set is very small.
Still, it is encouraging that all the objects except one have the
rotational periods which fall into the same color area as the
objects in our simulation (because of that we have not colored the
observed objects according to their period -- all except one would
be of the same color as the background). All of the observed
objects are, in our terminology, double asteroids. The single fast
rotators have not been quantitatively compared to the observed
objects (since there are no other dynamical parameters to compare
for single systems but the period of rotation) but their formation
is obviously reconstructed in the simulation.

Therefore, we can say we have succeeded to give possible
explanations of some typical evolution paths among NEA although we
failed to reproduce the single slow rotators. In the following
section, we shall try to give theoretical interpretation of these
results, and to discuss some other possible consequences.

\section{Discussion}

Results of our research, although speculative and tentative in
nature, do somewhat explain formation of some types of NEA. Our
basic concept -- simultaneous simulation of both orbital motion
and collision/disruption events -- seems to have shed some light
on the interrelations between these aspects of NEA evolution.
Namely, collision and disruption events, which lead to formation
of NEA, are strongly associated with transition mechanisms so it
seems unnecessary to introduce cosmogonic influences. Of course,
disadvantages of this concept are also clear -- large amount of
calculations, somewhat difficult precise interpretation of results
due to many factors and processes involved and uncertain initial
conditions for some parameters (e. g. spin vectors, shapes, etc).

Dynamical evolution of binaries shows some similarity with the
research of \cite{chau} although, as already mentioned, we have
not noticed importance of spin vector orientation. Detailed
treatment of internal heating in our model (Tillotson EOS) has
probably allowed us to get a more realistic picture of where the
energy of tides in retrograde systems goes: it becomes lost in
internal ``geological activity''.

Continual ejection of small fragments during collisions may be an
explanation for the overabundance of small Earth crossers, noticed
by some authors \cite{michcemda,rabin}. Again, cosmogonic
influences seem to be unnecessary if these fragments are taken
into account. This could also be a source of some meteor streams
(e. g. Geminids). The latter idea is, clearly, only a speculation.

Overall, tidal forces seem to play a more prominent role than
collisions. This is in part a consequence of more realistic
initial conditions for close approaches but it also seems that our
model treats tides better than collisions -- the latter require
better resolution, taking into account effects of fractures, etc.
It seems hard to estimate the consequences of the neglecting of
friction. Absence of friction limits the deformation an asteroid
can withstand with no disruption and therefore lessens the
magnitude of large deformations but, on the other side, it makes
small deformations easier. This is qualitatively similar to the
conclusions of \cite{solhil} and \cite{lovahrensb}.

The most intriguing result is, of course, the lack of single slow
rotators and the overabundance of asteroids with a satellite and
synchronized binaries. As already mentioned, the reason for the
latter lies partly in disadvantages of observational techniques.
However, the problem of single slow rotators must be connected
with some intrinsic disadvantages of the simulation. The formation
of slowly rotating systems is generally an open problem
\cite{pravhar,harris}. Spectrum of possible explanations
(concerning our simulation) is very diverse: from inadequate model
(equation of state, detection of satellites, etc) to inadequate
initial conditions (i. e. some kind of cosmogonic influences) to
short interval of simulation. There are some speculations
\cite{brunini} that various groups of asteroids may have different
cosmogonic origin and different early evolution paths. We have
also been suggested (\'{C}irkovi\'{c}, personal communication)
that slow rotators might be of cometary origin. In that case,
their former cometary activity might greatly influence their
present dynamical and physical state. Finally, bearing in mind the
specific optical properties of some observed slow rotators (e. g.
4179 Toutatis), we think that non-gravitational forces may also
have some influence on their spin vectors.

It is, however, more interesting to discuss the possible
interrelation between the lack of single slow rotators and the
overabundance of synchronized binaries. If the latter is not
completely originated in selection effects, it might be possible
that some synchronized systems can, over time scales longer than
our simulation, loose one of the components and become single
slowly rotating asteroids. Unfortunately, this does not seem too
realistic the framework of our model since we have not noticed any
such phenomenon although the time scale of our simulation is
approximately equal to the average lifetime of NEA. Therefore, one
has to assume that some more sophisticated processes, first of all
the non-gravitational effects (see \cite{pravhar} for a discussion
of possible influence of Yarkovsky force on the rotation of
asteroids; we think similar effects can influence also the
revolution of the components around the mutual center of mass )
play the key role in such events. So, we can conclude that some of
the synchronized binaries we predicted in the simulation do exist
but still escape the observational detection while the rest of
them actually dissolve into single fast rotators.

Many questions remain open. A more detailed physical model of
asteroids would allow a more refined treatment of
collision/disruption events. The primary task would be to include
friction and fractures. More realistic treatment of
non-gravitational forces (concerning the discussion in the
previous paragraph) could also give some new explanations. A
better description of chemical and elastic properties of the
asteroid material (basalt is only a phenomenological
approximation) is also one of possible enhancements for this kind
of research.

\section{Conclusions}

We have carried out a simultaneous numerical simulation of
migration and evolution of NEA. This has allowed us to investigate
the interrelation of orbital and physical evolution in a more
realistic way than in previous, isolated numerical researches. We
have confirmed a strong correlation between the formation of
binary systems (found to comprise about 12$\%$ of NEA population)
and events typical for the transition process which makes it
unnecessary to introduce cosmogonic influences in order to explain
the hypothesis, suggested by the observations, that an
overabundance of binaries among NEA exists in comparison to the
Main Belt.

We have detected formation of four typical products of tidal
and/or collisional evolution: double systems, synchronized
systems, asteroids with a satellite and single fast rotators. The
simulation gives, in our opinion, sufficiently robust and
realistic models of their formation. Asteroids with a satellite
and single fast rotators are formed in collisions, the primary
difference being the mass ratio of the colliding bodies: the
formation of the latter requires colliding bodies to be of the
same order of magnitude, and also some other conditions to be met
(e. g. impact angle, spin axis alignment, etc). Asteroids with a
satellite, however, seem to be very unstable, for unknown reasons.
This instability, together with observational selection effects,
can account for the lack of observational evidence of such objects
among the NEA.

Double asteroids (well known in observations) are a product of
tidal evolution. Synchronized binaries form either from the
previous type, or by tidal disruption. We failed to reproduce
extremely slow-rotating bodies, which could be due to their
peculiar cosmogonic origin or due to the disadvantages of our
simulation. We prefer the explanation that a fraction of
synchronized systems dissolve due to non-gravitational effects and
other sophisticated mechanisms not taken into account in this
simulation. The rest of synchronized binaries (those which
survive) are likely to be still undetected because of the
intrinsic limitations of current light curve inversion techniques.

Generally speaking, tidal forces have proven to be more important
than collisions. We noticed the continual formation of small
fragments in collision/disruption events, which can explain the
overabundance of these objects in the NEA belt and allow a
mechanism for keeping their population in a stationary state.

Of course, some results remain unexplained. Fate of synchronized
binaries and origin of single slow rotators remain puzzling.
Influences of friction and fractures, importance of
non-gravitational forces, etc. also require further numerical and
theoretical investigations.

\begin{ack}
The author wishes to thank to Zoran~Kne\v{z}evi\'{c} (Astronomical
Observatory, Belgrade) for helpful discussions and comments and
for provided references. Author also expresses his gratitude to
William~F.~Bottke (Department for Space Studies, Southwest
Research Institute, Boulder) for his valuable hints concerning the
hydrodynamical model (e-mail correspondence), and to Milan~
\'{C}irkovi\'{c} (Astronomical Observatory, Belgrade) for his
interesting suggestions about the origin of slow rotators. Thomas~
J.~Ahrens (Department for Geological and Planetary Sciences,
California Institute of Technology, Pasadena) was very kind to
provide the reference \cite{lovahrens}. Igor~Smoli\'{c}
(Department of Astronomy, Petnica Science Center, Valjevo)
indebted the author with careful reading of the paper and useful
comments.
\end{ack}

\newpage
\begin{table}
\caption{Basic properties of typical collisionally and/or tidally
evolved objects: physical origin, number of detected objects of
that type during the simulation, lifetime (in Myr), periods of
rotation and revolution around the center of mass, semimajor axis
of the secondary's orbit in units of primary's radii, eccentricity
of the secondary's orbit and ratio of the components' radii.
Types: 1
--- double asteroids, 2
--- synchronized binaries, 3 --- asteroids with a satellite, 4 ---
single fast rotators.\label{tab1}}
\begin{tabular}{|c|c|c|c|c|c|c|c|c|c|}
\hline Type &Origin &$N$ &$T$ [Myr] &$P_0$ [h] &$P$ [h] &$a$
[$R_p$] &$e$ &$R_s/R_p$ \\ \hline 1 &tides
&18 &1.1--4.9 &2--11 &14--36 &2.1--7.1 &0.02--0.58 &0.11--0.82 \\
\hline 2 &type 1 and tides &25 &1.4--7.3 &9--21 &$P\equiv P_0$
&0.0-7.9 &0.02--0.72 &0.19-0.89 \\ \hline 3 &collision &3
&0.0--1.2 &6--12 &17--33 &5.3--12.2 &0.15--0.48
&0.02-0.08 \\ \hline 4 &collision &9 &0.9--3.2 &2--4 &--- &--- &--- &--- \\
\hline
\end{tabular}
\end{table}

\clearpage
\begin{figure}[htbp]
\label{fig1} \caption{Spin rate histograms for simulated and
observed binaries. While the simulation gives a distribution
approaching to the Maxwellian, the observed objects show a
remarkable lack of slowly rotating binaries.}
\end{figure}

\begin{figure}[htbp]
\label{fig2} \caption{Relative abundaces of the four detected
types of evolved objects. Only single fast rotators and double
asteroids have been observed so far.}
\end{figure}

\begin{figure}[htbp]
\label{fig3} \caption{Contour plot of the period of rotation (for
binary systems which are not synchronized, period of the larger
component is given) upon the initial values of the semimajor axis
(in units of the primary's radius $R_p$) and the eccentricity of
the components. The darkest area (in the upper left part) in fact
corresponds to the pericentric distances smaller than the
primary's radius. Observed binary objects (the seven ones for
which an estimate of the eccentricity exists) are also given
(marked as stars).}
\end{figure}

\newpage
\begin{figure}[htbp]
\includegraphics[width=130mm]{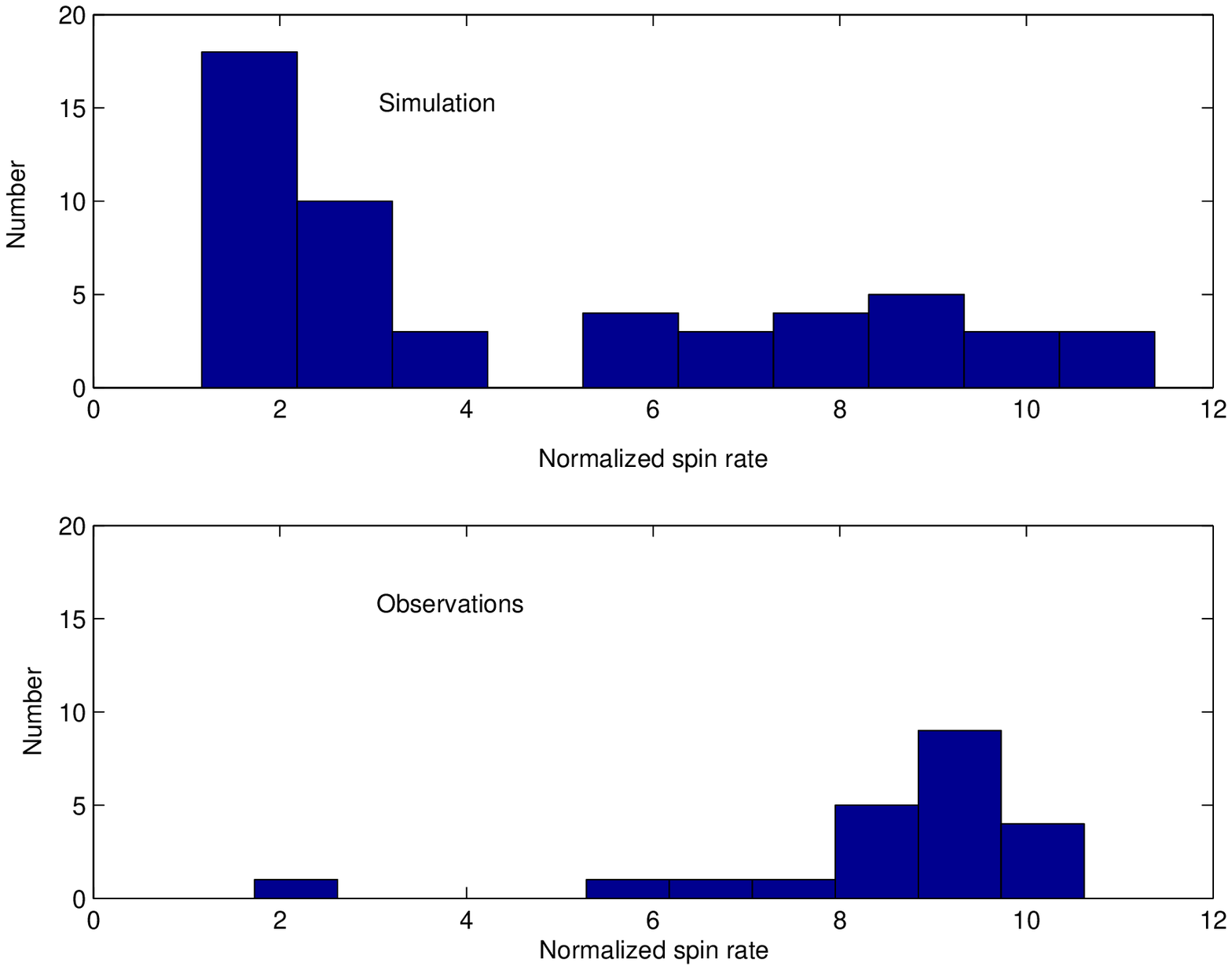}
\end{figure}
\center{Fig. 1}

\newpage
\begin{figure}[htbp]
\includegraphics[width=130mm]{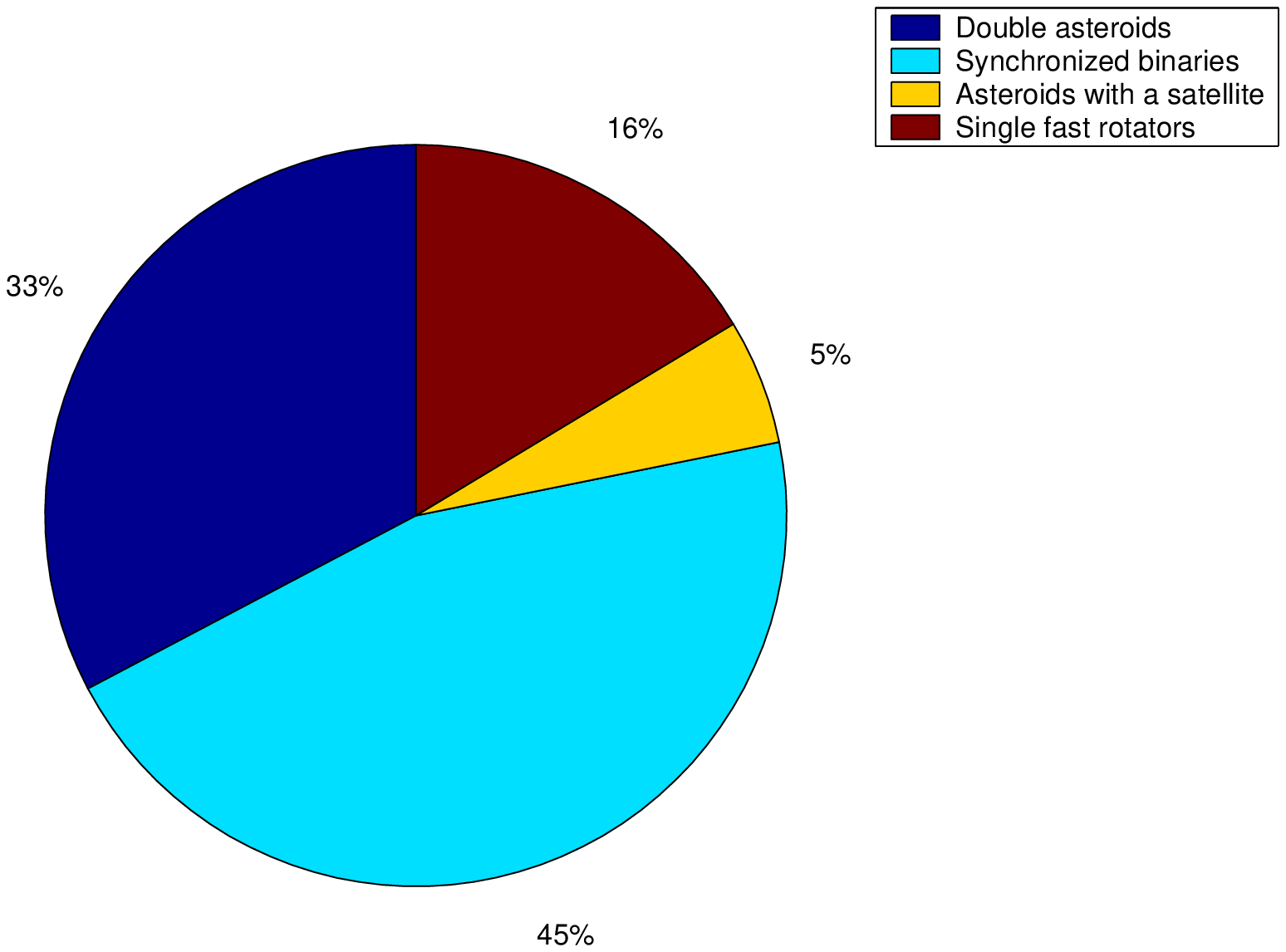}
\end{figure}
\center{Fig. 2}

\newpage
\begin{figure}[htbp]
\includegraphics[width=140mm]{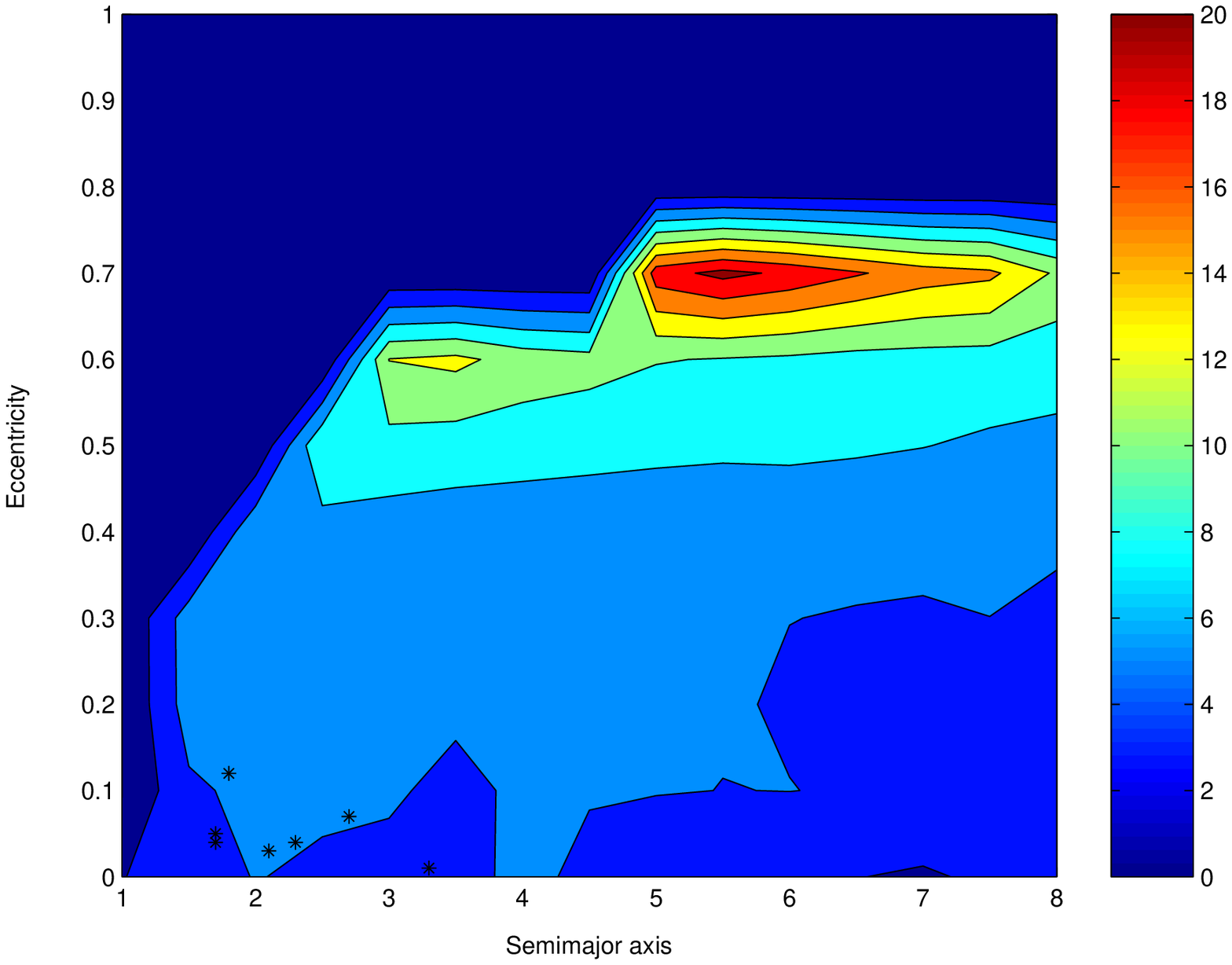}
\end{figure}
\center{Fig. 3}

\end{document}